# Comments on "Near-field interference for the unidirectional excitation of electromagnetic guided modes"

Seung-Yeol Lee,[1] Il-Min Lee,[1] Kyoung-Youm Kim,[2] and Byoungho Lee[1*]

[1]National Creative Research Center for Active Plasmonics Application Systems

Inter-University Semiconductor Research Center and School of Electrical Engineering

Seoul National University, Gwanak-Gu Gwanakro 1, Seoul 151-744, Korea

[2]Department of Optical Engineering, Sejong University, Seoul 143-747, Korea

* byoungho@snu.ac.kr

## Abstract

Rodríguez-Fortuño *et al.* (19 April 2013, p. 330) reported the unidirectional excitation of electromagnetic guided modes via the oblique illumination of a circularly polarized light. This comment points out that the same scheme was addressed in our a-year-ahead paper and that magnetic dipoles play a key role in the unidirectional excitation which was neglected in Rodríguez-Fortuño *et al.*'s report.

## Main text

Rodríguez-Fortuño *et al.* reported a way of unidirectional excitation of electromagnetic guided modes via near-field interference (*1*). This article was also explained in a Perspective (*2*). Their main concept was the near-field interference between horizontal and vertical dipoles that have phase retardation induced by illuminating circularly polarized light ($p_x$ and $p_z$ with $\pi/2$ phase difference). Following the numerical calculations to support their concept, experimental demonstration on the unidirectional excitation of surface plasmon polariton (SPP) waves from a nanoslit under the oblique illumination of elliptical or circularly polarized light was reported. Although their conceptual explanation via the near-field interference was interesting, the concept is neither very novel nor counterintuitive, as was claimed in their paper (*1*).

In a situation involving the excitation of a waveguide mode via a dipole source apart from the waveguide, provided that the mode index is larger than the effective index of the surrounding material (1.0 for air), the coupling to the guided mode will naturally be induced by the evanescent component of the dipole field. Since the guided modes have real momenta in their propagating directions, the directionality itself does not originate from the evanescent coupling but, rather, from the interference of the modes from two different dipole sources that have different symmetries. Therefore, in our opinion, nothing is counterintuitive in the concept reported by Rodríguez-Fortuño *et al*.

However, this is not the main reason for submitting this *Technical Comments*. In fact, the unidirectional excitation of SPPs with the same configuration as Fig. 3 of ref. (*1*) and Fig. 1C of ref. (*2*) was first reported with fully resolved physical explanations in a-year-ahead report by us (*3*). In that work, we not only implemented the unidirectional launching of SPP waves but also revealed the mechanism responsible for it: the asymmetric superposition of SPP waves generated by an electric dipole and a magnetic dipole.

In addition, we found that the experimental conditions of Rodríguez-Fortuño *et al*. do not constitute a valid support of their theoretical estimation. Their theory was based on the circularly polarized two-dimensional dipoles. To implement such a source, i.e., horizontal and vertical line dipole sources having π/2 phase shift, they adopted oblique illumination of a plane wave into a nanoslit. That is the experimental setup in Fig. 3 in ref. (*1*), where the oblique illumination seems to be devised because of the practical difficulties associated with implementing two-dimensional line dipole sources that are numerically treated in Fig. 2 of ref. (*1*). However, as was revealed in our earlier paper (*3*), the oblique incidence in Fig. 3 of ref. (*1*) is the key factor of the unidirectional excitation and thus should not be regarded as an approximation or practical expediency as in ref. (*1*). It induces the *y*-directional electric field component (parallel-to-slit), making the phase of the $p_x$ dipole oscillate along the slit. In this sense, along with the $p_x$ dipole, the magnetic dipole becomes another main contributor to the unidirectional excitation of SPP waves (*3*). This important contribution was not, unfortunately, pointed out in the work of Rodríguez-Fortuño *et al*. (*1*).

In Fig. 1, we present the field distribution radiated from line dipole sources placed on a metallic plane, calculated by the Green dyadic method (*4, 5*). It shows that the $p_y$ line dipole can also generate a surface mode when the phase is varying along the slit direction, and its coupling efficiency is in a similar order of magnitude. Therefore, a more comprehensible explanation would involve considering the composition of the anti-symmetric coupling of the surface mode caused by $p_x$ from *s*-polarization and the symmetrical coupling of the surface mode caused by $p_y$ and $p_z$ from *p*-polarization, or a *magnetic* dipole as in our work (*3*). In our analysis (*3*), the amount of induction currents near the slit edge was calculated and was expressed as magnetic dipole $m_x$, which takes into account both the *y*- and *z*-directional induction currents or the roles of $p_y$ and $p_z$. Therefore, the numerical results reported in ref. (*1*) do not fully explain the physical mechanism of experimental results in ref. (*1*), and a reasonable explanation was given in our previous work (*3*).

To strengthen our point, let us consider another example. In Fig. 2, we introduce a metal film to prove that the role of $p_y$ should not be neglected in the illumination conditions of ref. (*1*). This metal film can then guide coupled SPP modes between the upper and lower surfaces, which are often referred to as insulator-metal-insulator (IMI) modes. It is well known that there are two types of IMI modes: one being the long-range surface plasmon (LRSP) mode and the other the short-range surface plasmon (SRSP) one (*6*). The charges of the upper- and lower-side of the metal film for the LRSP have opposite signs, whereas those for the SRSP have the same signs. When the thickness of the metal film is far thinner than the incident wavelength, only the SRSP is excited in the case of the *s*-polarization incidence as shown in Fig. 2B ($p_x$). On the other hand, the dipole $p_z$ can be attributed to the charges in opposite signs, so only a LRSP is generated. Results shown in Fig. 2C indicate that both $p_y$ and $p_z$ contribute to the excitation of SPP with similar order of magnitudes. Moreover, the unidirectional launching of the IMI mode cannot be achieved by interfering $p_x$ and $p_z$ in this situation since they excite different types of modes, whereas that of $p_x$ and $p_y$ results in the unidirectional launching. Although the excitations from $p_y$ and $p_z$ dipoles are not distinguishable in the asymmetric surrounding configuration, we conclude that it is not sufficient to model the experimental results of ref. (*1*) without taking the parallel-to-slit directional oscillation of surface currents into consideration.

In short, we believe that the unidirectional excitation of SPPs from a slit as shown in Fig. 3 in ref. (*1*) and Fig. 1C of ref. (*2*) is not a new finding and such excitation should be understood by considering the interference of the $p_x$ dipole and the $m_x$ dipole as in (*3*).

**References**

1. F. J. Rodríguez-Fortuño, G. Marino, P. Ginzburg, D. O'Connor, A. Martínez, G. A. Wurtz, A. V. Zayats, *Science* **340**, 328 (2013).
2. A. E. Miroshinichenko and Y. S. Kivshar, *Science* **340**, 283 (2013).
3. S.-Y. Lee, I.-M. Lee, J. Park, S. Oh, W. Lee, K.-Y. Kim, and B. Lee, *Phys. Rev. Lett.* **108**, 213907 (2012).
4. L. Novotny and B. Hetch, *Principles of Nano-Optics* (Cambridge University Press, New York, 2006).
5. M. Paulus and O. J. F. Martin, *Phys. Rev. E* **63**, 066615 (2001).
6. H. Raether, *Surface Plasmons on Smooth and Rough Surfaces and on Gratings* (Springer, Berlin, 1988).

**Figure captions**

Fig. 1. The distribution of a *z*-directional electric field radiated by line dipole sources laid on a metal substrate is shown. Radiations from (A) $p_x$, (B) $p_y$, and (C) $p_z$ dipole line sources which have a sinusoidally varying phase profile along the line axis at $\lambda$ = 633 nm are calculated by the Green dyadic method to show the symmetry of the coupled SPP mode. The period of sinusoidal oscillation along the line axis is set to $\lambda_y = \lambda/\sin(70°)$, to follow the experimental conditions of ref. (*1*).

Fig. 2. (A) The type and symmetry of coupled IMI plasmonic modes for different orientation of line dipole sources are depicted. Due to the charge symmetry of the upper and lower metal surfaces, the short-ranged IMI mode is generated by $p_x$ and $p_y$, whereas the long-ranged mode is generated by $p_z$. Surface charges can be induced by a $p_y$ line dipole, since we assumed the oblique incidence case. (B) *z*-directional electric field under *s*-polarization and (C) absolute value of the *z*-directional electric field under *p*-polarization illumination on the floated metal film are shown. All other conditions follow the experimental condition reported in ref. (*1*). The incident wave, the directly reflected wave without scattering, and tunneled wave through thin metal film are not drawn to clearly show the SPPs.

Fig. 1

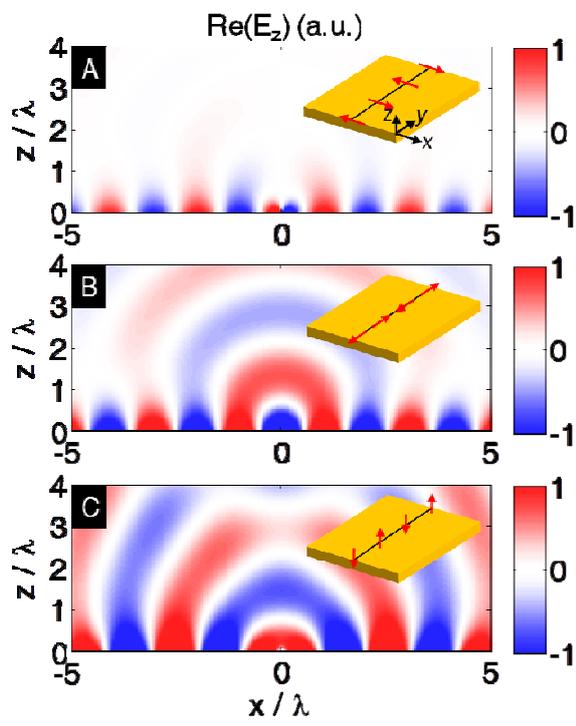

Fig. 2

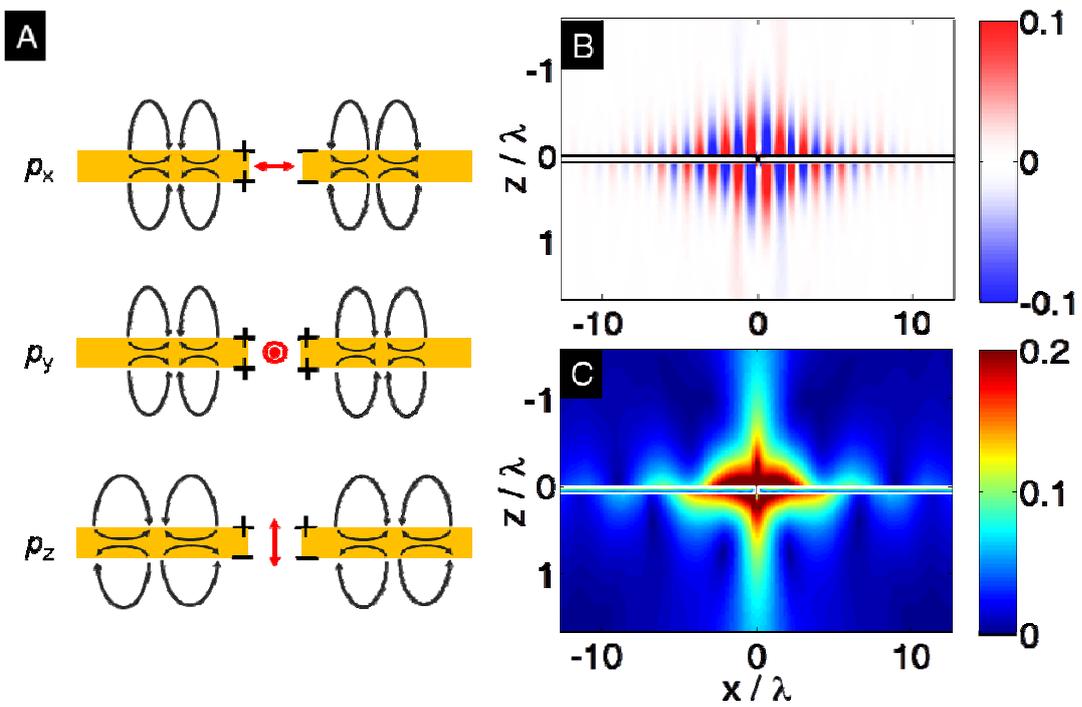